# A Case Study of the Modified Hirsch Index $h_\text{m}$ Accounting for Multiple Co-authors

Michael Schreiber

*Institut für Physik, Technische Universität Chemnitz, 09107 Chemnitz, Germany*

*E-mail: schreiber@physik.tu-chemnitz.de*

**Abstract.** – J. E. Hirsch (2005) introduced the $h$-index to quantify an individual's scientific research output by the largest number $h$ of a scientist's papers, that received at least $h$ citations. This so-called Hirsch index can be easily modified to take multiple co-authorship into account by counting the papers fractionally according to (the inverse of) the number of authors. I have worked out 26 empirical cases of physicists to illustrate the effect of this modification. Although the correlation between the original and the modified Hirsch index is relatively strong, the arrangement of the datasets is significantly different depending on whether they are put into order according to the values of either the original or the modified index.

**Introduction**

The *h*-index has been proposed by Hirsch (2005) as an easily determinable estimate of the impact of a scientist's cumulative research contribution. It is defined as the highest number of papers of a scientist that have been cited *h* or more times. Due to its simplicity it has soon become famous and attracted attention, which can be quantified by the high number of more than 70 citations which the article (Hirsch, 2005) accumulated within less than 2 years, thus enhancing Hirsch's Hirsch index. Of course, it is always dangerous to reduce the complete scientific output of the researcher to a single number. Nevertheless, it is a fact that the *h*-index is more and more utilized to quantify the visibility, importance, significance, and broad impact of individual scientists, departments, countries, or research fields. Its calculation has even been implemented already in the ISI Web of Science (WoS) provided by Thomson Scientific. However, this automatic determination has to be taken with considerable caution, because an undiscriminating WoS search often leads to a completely wrong data base usually comprising too many papers due to homographs, but sometimes too few papers due to misspellings, transliterated names, name changes, or similar difficulties. This is called the precision problem, which is well-known in bibliometrics and scientometrics and has been noted in numerous studies. For the present data set it has been discussed in detail previously (Schreiber, 2007a).

Advantages and disadvantages of the *h*-index have been discussed by many authors. Less than one year after the introduction of the *h*-index, already a review of the research literature on this topic was published (Bornmann & Daniel, 2007). In the meantime several variants of the *h*-index have been introduced, e.g., the *g*-index which is sensitive to one or several outstandingly highly cited manuscripts (Egghe, 2006), the *A*-index and the *R*-index which measure the citation intensity in the *h*-core (i.e., the *h*-defining set of papers) (Jin et al., 2007), or the $h_s$-index which takes self-citation corrections into account (Schreiber, 2007b). Recently nine different variants of the $h$-index have been compared (Bornmann et al., 2008).

One disadvantage of the $h$-index is its insensitivity to the numbers of co-authors of a given publication. This was already noted by Hirsch (2005). He proposed "to normalize $h$ by a factor that reflects the average number of co-authors". This normalization has been applied (Batista et al., 2006) using the mean number of authors of the papers in the $h$-core for the normalization and the resulting index has been labelled $h_I$. The authors of that study, however, have already noted that the average is sensitive to extreme values. This means that the influence of single-author publications to one's $h$-index can be strongly reduced. On the other hand a few papers with a large number of co-authors will lead to an excessively large normalization (Schreiber, 2008a).

In scientometrics, the problem of how to count multi-authored publications has been discussed for a long time (Lindsey, 1980, Price, 1981), assigning credit proportionally to the number of authors which is usually called fractional counting or adjusted counting. There have, however, evolved a number of different methods for accrediting publications for several authors, see e.g. Egghe et al. (2000). It is widely accepted that some kind of discounting should be applied (Harsanyi, 1993, Lukovits & Vinkler, 1995, Burrell & Rousseau 1995, Egghe et al., 2000, Trueba & Guerrero, 2004) also to the Hirsch index (Jin et al., 2007, Burrell, 2007, Wan et al., 2007, Egghe 2008). One difficulty is, that different scoring methods can lead to paradoxical effects and yield totally different rankings (van Hooydonk, 1997, Egghe et al., 2000) so that no unambiguous solution of the "multiple-author problem" (Harsanyi, 1993) exists. But fractional counting is usually preferred since it does not increase the total weight of a single paper (Egghe et al., 2000). Egghe and Rousseau (1990) stated already "that the best way to handle multi-authored papers is to assign credit proportionally."

I have recently proposed (Schreiber, 2008a) to modify the $h$-index by counting the papers fractionally according to (the inverse of) the number of authors yielding the modified index $h_m$. Analysing the citation records of 8 famous physicists it was shown that this can have a significant influence and lead to a different ranking than the original $h$-index. The same fractional counting of papers has been suggested by Egghe (2008) and applied to two fictitious examples and one empirical case. But the effect was relatively small, because of a large number of single-author papers in his data set. It is the aim of the present paper to demonstrate the effect of the fractionalised counting on the citation records of 26 not-so-prominent physicists. Thus the obtained observations should be more common for the data sets of more average scientists.

The validity of a new index should be analysed on the basis of empirical data. This is the purpose of the present investigation, which is restricted to the comparison of the original Hirsch index, the normalized index $h_I$ and the modified index $h_m$. In principle, most other ways which are available in the literature to distribute credit among co-authors could be applied to the Hirsch index as well. But these ways are usually more complicated and require assumptions about the relative contribution of different authors. Without specific information about, e.g., a particularly high contribution of the first or the last author in the author list, I believe that the fairest way of attributing the credit is to share it among all co-authors as it is applied in the following.

**The Data Base, the Computation of the Modified Index $h_m$ and its Visualization**

The data for the subsequent analysis have been compiled in January and February 2007 from the Science Citation Index provided by Thomson Scientific in the WoS. The 26 citation records have been analysed with respect to the self-citations (Schreiber, 2007a). As specified in that publication, the 26 data sets include the records of all full and associate professors from the Institute of Physics at my university, some recently retired colleagues, as well as all scientists that have been working as assistants or senior assistants in my group doing their research for their habilitation degree or afterwards. The data sets are labelled A,B,C,…,Z in conformity with the previous analysis (Schreiber, 2007a).

The same data have been utilized for an investigation of the *g*-index in comparison with the *h*-index, the *A*-index and the *R*-index (Schreiber, 2008b). As mentioned in the introduction, the precision problem means that a simple WoS search is not sufficient. For the present investigation, reasonably great care has been taken to establish the correct data base. As detailed before (Schreiber 2007a) homographs yield an enhancement of the *h*-index in 9 cases, in 6 cases this was 50% or more, in one case even a factor of 2.73. For two data sets the reverse problem was encountered, because important publications were missed by the general search in the ISI WoS.

The WoS allows an automatic arrangement of the publication lists in decreasing order according to the number of citations $c(r)$, where $r$ is the rank attributed to the paper. The *h*-index is readily read off this list as

$$c(h) \geq h \geq c(h+1) \qquad (1)$$

according to Hirsch's original definition (Hirsch 2005). It is worth noting that several people believe that the original definition demands that the papers beyond the rank *h* should "have fewer than *h* citations each". This wording was indeed chosen by Hirsch in the first preprints. However, this formulation would make the index not quite well-defined in all cases. But in the last version available on the preprint server arXiv:physics as well as in the final publication (Hirsch 2005) the phrase has been corrected, requiring now that the papers beyond the rank $h$ should "have $\leq h$ citations each", with which the second inequality in Eq. (1) is in conformance.

In Eq. (1) each paper is fully counted for the (trivial) determination of its rank

$$r = \sum_{r'=1}^{r} 1.  \qquad (2)$$

The upper histogram in Figure 1 shows the citations arranged in this way for data set C. The intersection with the white line, which displays the function

$$c(r) = r,  \qquad (3)$$

yields the $h$-index. For the data set in Figure 1 one obtains $h^C = 23$.

If one counts a paper with $a(r)$ authors only fractionally, i.e., by $1/a(r)$ one obtains an effective rank

$$r_{\text{eff}}(r) = \sum_{r'=1}^{r} \frac{1}{a(r')}  \qquad (4)$$

This can be utilized to define the modified index $h_m$ as

$$c(r(h_m)) \geq h_m \geq c(r(h_m)+1)  \qquad (5)$$

where $r(h_m)$ follows from the inverse function $r(r_{\text{eff}})$ of Equation (4). This means that $h_m$ is that effective number of papers which have been cited $h_m$ or more times, while the further papers have no more than $h_m$ citations each. It is easy to visualize this definition by plotting the respective histogram with bar widths which are determined by (the inverse of) the number of authors for each paper, as shown in the middle histogram in Figure 1. This leads to a significant compression of the histogram towards lower values of the rank. Correspondingly the effective number of papers in the $h$-core is much smaller than $h$, for the data in Figure 1 one obtains $r_{\text{eff}}^C(h^C) = 6.33$.

Therefore beyond $h^C$ there are papers with more citations than this effective rank. These have to be taken into account for the modified index, as visualized in Figure 1. It is indeed nearly always a considerable number of publications with citation counts between $h$ and $h_m$ which contribute to the $h_m$-core, i.e., to the $h_m$-defining set. In the considered example, the $h_m^C$-core comprises $r(h_m^C) = 41$ publications with an effective rank $h_m^C = r_{\text{eff}}^C(41) = 11.03$. This value can also be read off the intersection of the function $c(r) = r$ with the middle histogram in Figure 1. It means that there are 41 publications with at least 12 citations in the $h_m^C$-core. The 42$^{nd}$ paper attracted only 11 citations, in agreement with Equation (5).

For a large data set the visualization of the citation records as in Figure 1 is probably the easiest way to present the data and to allow an assessment of the influence of multiple co-authors. But in order to demonstrate the calculation of the individual indexes in more detail, the citation records of 4 scientists with a relatively small number of publication are presented in Table 1. There only the 20 most cited publications are included for each data set. These are sufficient for determining not only the original Hirsch index, but also the modified index. The original Hirsch index is easily read off this table as the highest rank $r$ for which the citation count $c(r)$ is larger than or equal to the rank. The values of the effective rank $r_{\text{eff}}(r)$ in this table are determined according to Equation (4), i.e. counting the papers fractionally according to the number of authors for each paper. Consequently, the modified index can also be easily read off this table as the largest effective rank for which the citation count is larger than or equal to this effective rank.

The 4 data sets in Table 1 represent quite different citation records, data set X is characterized by a very high citation count of the first publication, and also in case V $c(1)$ is quite large. Data sets V and W show a large tail, i.e. the number of citations for the last papers in the table is relatively high which is reflected in the observation that 6 and 9 papers enter the $h_m$-core in addition to the 10 and 9 papers in the $h$-core, respectively. For data set X the number of co-authors is quite large for all publications, yielding of course small effective ranks which in turn leads to a large increase of the size of the $h_m^X$-core from $h^X = 8$ to 15 papers, which is, however accompanied by a relatively small modified index, because the citation counts drop quite strongly in this range. An even stronger decrease of the citation counts can be observed for data set Y, but in this case the number of co-authors is relatively small (on average less than two), so that only one more paper contributes to the $h_m^Y$-core in addition to the 7 papers in the $h^Y$-core. It is therefore not surprising that the resulting modified index $h_m^Y = 4.83$ is significantly larger than $h_m^X = 2.95$ and also larger than $h_m^W = 4.33$.

**The Computation of the Normalized Index $h_I$ and its Visualization**

For the simple normalization of the $h$-index, the average number of authors of the first $r$ papers is calculated as the mean

$$\overline{a}(r) = \frac{1}{r}\sum_{r'=1}^{r} a(r') \qquad (6)$$

and utilized to determine

$$h_I = \frac{h}{\overline{a}(h)}. \qquad (7)$$

If one employs Equation (6) for the (trivial) determination of a normalized rank

$$r_I(r) = \frac{1}{\overline{a}(r)}\sum_{r'=1}^{r} 1 = \frac{r}{\overline{a}(r)} \qquad (8)$$

and utilizes respective normalized citation counts

$$c_I(r) = \frac{c(r)}{\overline{a}(h)} \qquad (9)$$

then one can determine the normalized index $h_I$ in analogy to Equation (5) from the inequalities

$$c_I(r(h_I)) \geq h_I \geq c_I(r(h_I)+1) \quad (10)$$

where $r(h_I)$ follows from the inverse function $r(r_I)$ of Equation (8) and we have $r(h_I) = h$. Of course, this complicated calculation is not necessary, because the simple definition (Equation 7) is sufficient. But I have made this detour in order to show that the straightforward normalization of the h-index not only means a scaling of the ranks (Equation 8) similar to Equation (4) but also a scaling of the citation counts (Equation 9). This can be easy visualized, compare Figure 1 where the lowest histogram is compressed towards the left by the factor $1/\bar{a}^C(h) = 1/4.22 = 0.24$ as well as downwards with the same factor. This factor yields the normalized index $h_I^C = 5.45$ which can also be read off Figure 1 from the intersection of the function $c(r) = r$ with the lower histogram. Effectively thus the calculation of the $h_I$-index means a fractionalised counting of the citations as well as a fractionalised counting of the papers, in each case by the mean number of authors. Consequently the impact is dramatically reduced as noted by Vinkler (2007). This double normalization is at least questionable.

**Results for the different indices**

The citation counts for 5 further data sets are presented in Figure 2 and the determination of the indices is visualized.

For data set D it is conspicuous that no single-author publications show up in the h- and not even in the $h_m$-core, as can be seen in the middle histogram, in which all respective bars are compressed. (In fact the first single-author paper appears at $r = 113$ with $r_{eff} = 21.75$ reflected by the wide white bar in the middle histogram at this rank.) This leads to a particularly strong normalization with $\bar{a}(h) = 6.05$, reducing the index from $h^D=20$ to $h_I^D = 3.31$. But in comparison with the case C in Figure 1, there are many papers with citation counts below but close to $h^D$. Therefore the upper histogram is rather flat beyond this value, which is of course reflected in the likewise quite flat middle histogram for the fractional counting of papers. Consequently $h_m^D = 10.97$ is very close to $h_m^C$.

In contrast, data set E comprises a large number of highly cited single-author publications, so that the histogram for fractional counting remains close to the full histogram at least for small ranks. It is thus not surprising that for both variants of the $h$-index the resulting values for this data set are higher than for data sets C and D, even though the citation counts decrease quite strongly beyond $h^E$.

The average number of co-authors in the $h$-core for data sets N and P is about the same as for E, so that the reduction between $h$ and $h_I$ is comparable. But the effect on the modified index is quite different. Due to the strong decrease of the citation counts $c^P(r)$ the $h_m^P$-core includes only one more paper than the $h^P$-core, leading to a relatively small value of $h_m^P = 6.92$. In contrast, 7 papers, three of them single-author publications enter the $h_m^N$-core in addition to the 14 papers in the $h^N$-core. Accordingly, the resulting value for $h_m^N = 11.50$ is relatively large. These differences are visualized in Figure 2 by the different widths of the central section in the middle histograms, compare case P with N.

A similar observation can be made comparing the central sections of the middle histograms in Figure 2 for cases E and D. With 23 publications the $h_m^E$-core comprises only 4 more papers than the $h^E$-core, while for the $h_m^D$-core 37 papers have to be added to the 20 publications in the $h^D$-core. But this difference is counterbalanced by the large number of single-author publications of author E, so that $h_m^E > h_m^D$ as mentioned above.

An extreme case occurs for data set G, because in this case 15 out of 17 publications in the $h^G$-core are single-author papers. Consequently, the upper histogram can barely be distinguished from the middle histogram. Moreover, there is no central section in the middle histogram, because the $h_m$-core comprises the same publications as the $h$-core. Nevertheless, the reduction to $h_I^G = 10.3$ visualized by the compression in the lower histogram is quite strong, because one paper with 11 authors has a strong effect on $\bar{a}(h)$. This is an example for the above mentioned fact that the average is sensitive to extreme values.

**The effect of the fractionalised counting on the ranking**

The resulting values of the indices for the 26 data sets are compiled in Table 2. In Figure 3 the obtained indices are displayed on a logarithmic scale so that the relative changes can be easily visualized.

It is obvious that a strong reduction of the Hirsch index occurs, when the number of co-authors is taken into consideration as quantified by the ratio $h_m/h$, which is on average $\langle h_m/h \rangle = 0.58 \pm 0.13$. Only for the data set G the effect is very small, as anticipated from the above discussion. It is interesting to note that in this case the omission of the 11-author publication would result in expectable reductions to $\overline{h^G} = 16$ and $\overline{h_m^G} = 15.5$ (no other papers enter the $\overline{h^G}$- or the $\overline{h_m^G}$-core), but also in a surprising <u>increase</u> to $\overline{h_I^G} = 15.06$ because the mean number of authors decreases strongly to $\bar{a}(h) = 1.06$. This is a very strange effect: neglecting a highly cited paper, in this case with $c^G(4) = 34$, leads to a significantly higher $h_I$-index. For the original index such a strange behaviour cannot occur nor is it possible for the modified index $h_m$. This is certainly an extreme case. But it is not unique. A closer inspection of the data for case X in Table 1 shows that deleting the most cited paper (which incidentally has 8 authors) from this data set would lead to a similar effect: in this case the Hirsch index would not change, because another paper with $c(8) = 8$ citations enters the $h^X$-core, but this has only five authors, so that the mean number of authors in the core decreases to $\overline{a}(h) = 4.88$. As a consequence, the normalized index increases to $h_I^X = 1.64$, that is the same strange behaviour as for the above example in the case G. On the other hand as expected the modified index decreases somewhat, in this case to $h_m^{\overline{X}} = 2.86$.

The relatively small changes for data set N have also been discussed in the previous section already. Likewise, data set Y is characterized by a very small average number of co-authors and a corresponding small effect for $h_I^Y$. But the reduction of the modified index as compared to the original index is not so small, because again a strongly decreasing citation record beyond $h^Y = 7$ allows only one publication to enter the $h_m^Y$-core in addition to the $h^Y$-core, as discussed above.

The values $r_{\text{eff}}(h)$ in Table 2 reflect, how strongly the h-core is compressed by the fractionalised counting of the papers. Of course, $r_{\text{eff}} > h_I$ in all cases, because the average (Equation 4) of the inverse numbers of authors is always larger than the inverse of the average (Equation 6), see also Figures 1 and 2.

The values $r(h_m)$ which are also given in Table 2, show how many papers contribute to the $h_m$-core and thus demonstrate how many more papers beyond the h-core have to be taken into account. Here the strong or weak decrease of the citation records in dependence on the rank as discussed above is significant. Comparing $r(h_m)$ with $h$ gives an indication, how much more severe the precision problem becomes for the determination of the modified index, because a significantly larger number of publications has to be evaluated.

The last column in Table 2 shows the rank which the data sets would hold after arranging them according to the modified index. Of course, the smaller changes should not be overinterpreted, but there are some very large rearranging effects, e.g. the colleagues N and Q move forward 9 and 8 places, respectively, in the $h_m$-sorted list. On the other hand the scientists H and J fall back 13 and 12 places, respectively. These rearrangements are illustrated in Figure 3, where the $h$-values yield relatively low bars for the cases N and Q in the left part of the diagram and relatively high bars for the cases H and J in the right part.

The observations can be quantified by calculating Spearman's rank-order correlation coefficient. The value $\kappa(h,h_m) = 0.755$ shows that there is of course quite some correlation which is not surprising, because the ratio $h_m/h$ is not so different for most data sets. The correlation $\kappa(h,h_I) = 0.597$ is significantly weaker. It is interesting to note that the correlation $\kappa(h_m,h_I) = 0.795$ is strongest. It is not unexpected that such significant correlations between the indices exist, because the consideration of the multiple co-authorship does of course not lead to a completely different valuation of the citation records. That means, that the indices might be called redundant and one might be tempted to consider just the weakest correlation as an incremental contribution of the $h_I$-index compared with the $h$-index. However, in my opinion it is important that some kind of discounting is only fair especially when a large number of co-authors has contributed to a publication. Therefore it appears to be appropriate to modify the Hirsch index, although the modified index is strongly correlated with the original index. It is interesting to note that the correlation between the $h$-index and other variants like the $g$-index and the $R$-index for the same data sets as in the present investigation are much higher (Schreiber, 2008b). The above given value $\kappa(h,h_m)$ is of the same order as the correlation between the Hirsch index and the total number of papers or the total number of cited papers (Schreiber, 2008b).

**Conclusions**

The modified Hirsch index $h_m$ was introduced (Schreiber, 2008a) to account for multiple co-authors in a reasonable way. It stands to reason that it is necessary to test the validity of a new index thoroughly on the basis of empirical data, before using such an index for comparison. The present case study provides such an analysis. Of course, the values of the modified index are always smaller than the original index, and the reduction depends on the number of co-authors. Naturally, scientists with many single-author publications become more prominent in the $h_m$-sorted list. This is the main effect of the modification. But the individual citation records can also have a strong influence on the modified index because they determine how many more papers contribute to the $h_m$-core in addition to the $h$-core. Authors with a rather flat frequency function $c(r)$ of citations are favoured by the modified index, which in my opinion is appropriate.

Of course, the precision problem increases for the determination of the modified index in comparison with the calculation of the original Hirsch index, because more papers have to be taken into account. This might be considered a disadvantage in contrast to the simple normalization of the $h$-index with the mean number of authors. However, that normalization appears to be unreasonably strong, because it effectively means that not only the papers are fractionally counted, but also the citations are fractionalised. The strange effect which a single paper with a large number of authors can have on the normalized index $h_I$ as observed above for data sets G and X, can also be elicited in the following reversed way (Schreiber, 2008c): if a publication with many authors enters the $h$-core, because its number of citations has increased, then this can lead to a decrease of the $h_I$-index; which is certainly a peculiar behaviour for an index that is supposed to measure the impact of the publications in terms of the numbers of citations. For the modified index such a problem does not occur.

The fractional counting of papers is not the only way to allocate the credit to several authors of a manuscript. Another straightforward fractional crediting system is to divide the number of citations by the number of authors for each paper (Egghe, 2008). In this way the citations are shared between the co-authors. However, in order to determine a Hirsch-type index this counting requires that the citation records are rearranged according to the fractional citation counts. Moreover, it leads to a fundamental problem when data sets are aggregated, e.g. when one determines the combined index of several people like all scientists in an institute. For example, a publication with two authors from that institute would contribute two times one half to their $h_m$-indices, if the citation count is large. But it would be fully taken into account twice when the citations are fractionalised, provided that the number of citations is sufficiently large. This is a methodological problem, which also occurs for the original Hirsch index $h$, as well as for the normalized index $h_I$. In contrast, the total weight is preserved for the modified index as it should (Egghe et al., 2000).

In summary, I think that there are three reasons because of which the modified index $h_m$ is more appropriate than the $h_I$-index, i.e. the straightforward normalisation of the original Hirsch index $h$ by the mean number of co-authors in the $h$-core: (i) For $h_I$ not only the fractional paper count but also the fractional citation count is utilized, which yields a double normalization and is thus excessive. For the modified index, only the rank is fractionalised. (ii) The total weight of a publication is preserved in the aggregation of citation records for different data sets, e.g. when combining the citation records of several scientists in an institute or of several institutes in a university, or of several universities in a country. In contrast, the weight is not preserved for $h_I$. (iii) Adding a highly cited publication to a data set always increases the modified index, but it can decrease the $h_I$-index, if the number of authors is above average. In conclusion there are at least 3 methodological advantages of the modified index $h_m$ in comparison with the normalized index $h_I$.

More complicated countings have also been proposed (Egghe et al., 2000, Wan et al. 2007) taking into account that the order of the names in the author list might give an indication about their relative contributions to the whole research. Whether this is indeed the case, depends on the customs in the field, and may also be different for different groups in the same field. Therefore in my opinion it is most appropriate to share the credit equally among all authors, as long as one does not have good information about their relative contributions. Ideally, information obtained directly from the authors about their individual contributions should be used for determining the impact (Vinkler, 1993).

From the point of view of a scientist who cites a publication, the number of co-authors of that publication is usually irrelevant. Accordingly the value of a citation and thus the impact of a paper should be taken into account independently of the number of authors. This is exactly what happens in the calculation of the modified index, because as discussed above the total weight of a publication and thus the value of its citation is preserved for the modified index.

Concluding, I believe that the modified index is the fairest way of taking multiple authorship appropriately into consideration if one attempts to quantify the impact of a scientist's cumulative research contribution (Hirsch, 2005) by a single number. Whether such an assessment is reasonable on the whole, remains a matter of controversy and I close with a word of caution (which has been attributed to A. Einstein): "Not everything that counts can be counted. And not everything that can be counted counts."

FIG. 1. The citation counts for the papers in data set C. The upper histogram with wide bars shows the numbers of citations $c(r)$ versus the rank $r$ which is attributed to each paper by sorting according to $c(r)$, up to the $h$-index (red/dark grey) and beyond (orange/medium grey). In the middle histogram the effective rank is used so that the original histogram is compressed towards the left (yellow/light grey for the first $h$ papers, turquoise/medium grey up to the $h_m$-th paper and white beyond). In the lower histogram the normalization with the mean number of authors of the first $h$ papers is used, so that the original histogram is compressed to the left as well as downwards (light green/medium grey up to the $h_I$-index and dark green/dark grey beyond). Note the logarithmic scale for $c(r)$. The thick white line displays the function $c(r)=r$, so that its intersections with the histograms (from top to bottom) yield the $h$-index, the $h_m$-index, and the $h_I$-index, respectively.

FIG. 2. Same as Figure 1, but for 5 different data sets D, E, G, N, and P.

FIG. 3. Hirsch indices for the 26 investigated data sets. From top to bottom: original Hirsch index $h$ (dark grey/red), modified index $h_m$ due to fractional counting of the papers according to the number of authors (light grey/yellow), and index $h_I$ determined by normalization of the $h$-index with the mean number of authors (green/medium grey). The data sets are put into order according to the modified index $h_m$, as indicated at the horizontal axis, where the letters are not in alphabetical order in contrast to the sequence in Table 2 determined by the original index $h$. Note the logarithmic scale for the $h$-values.

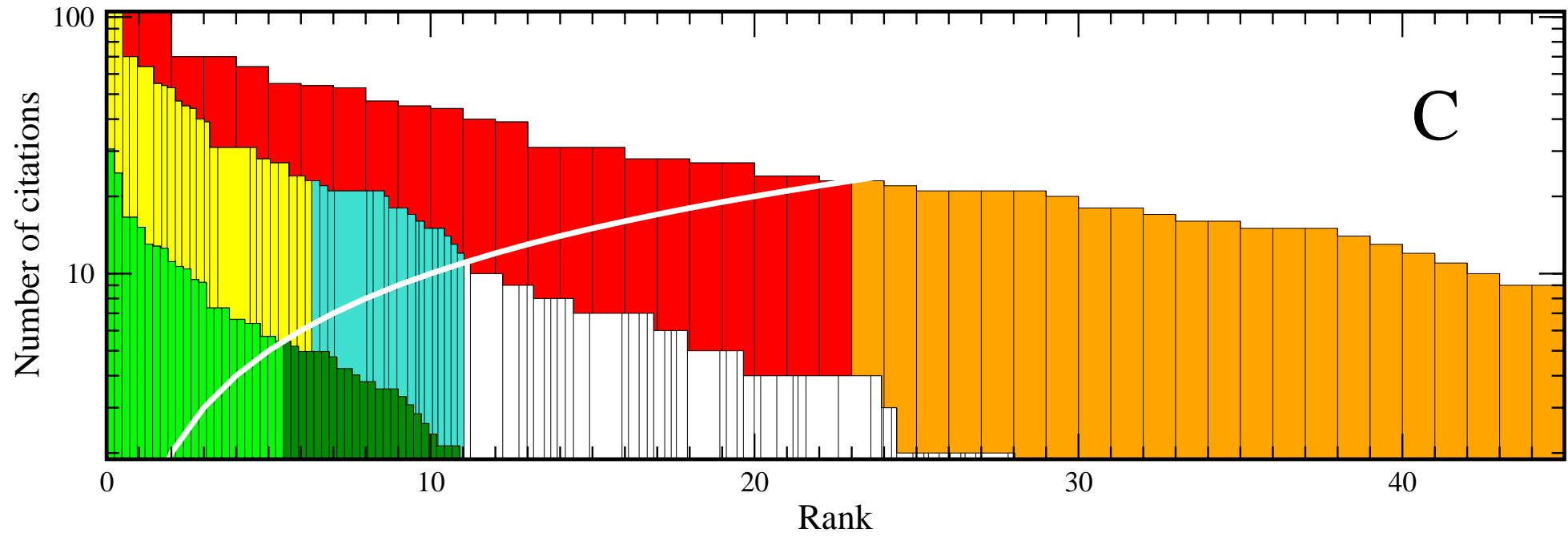

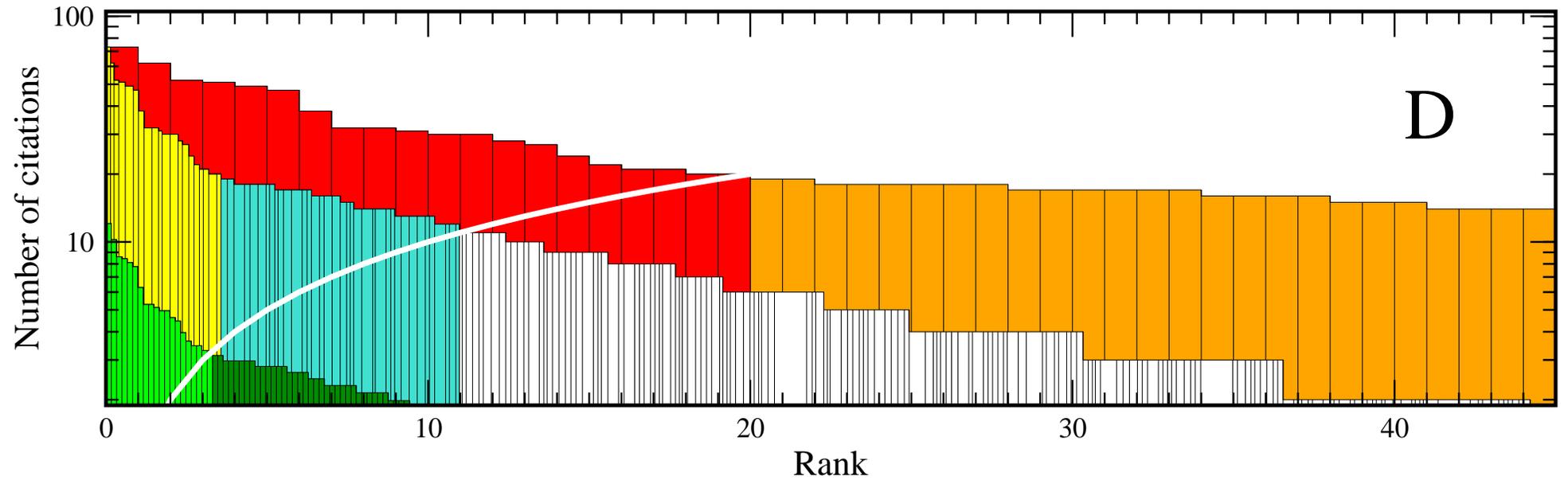

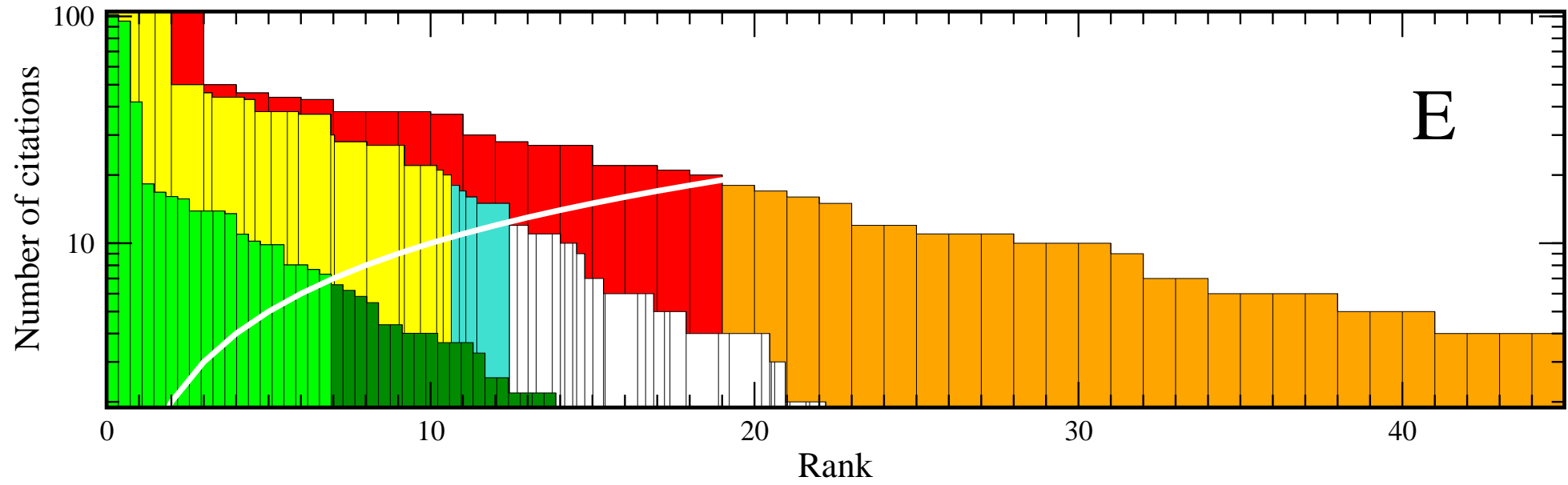

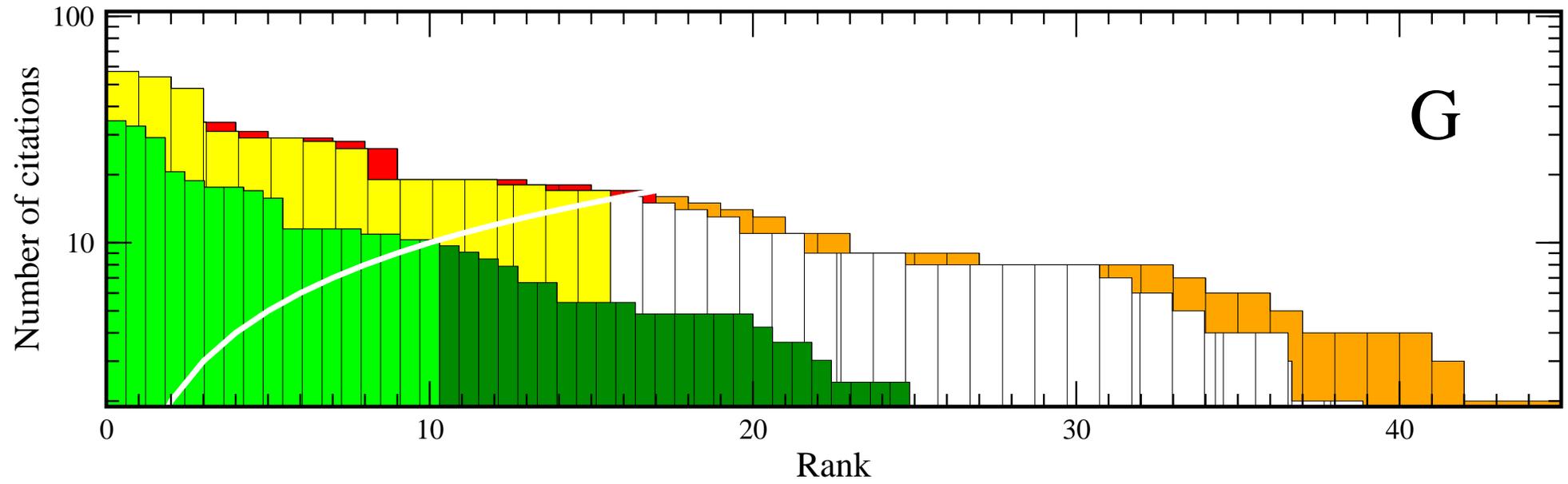

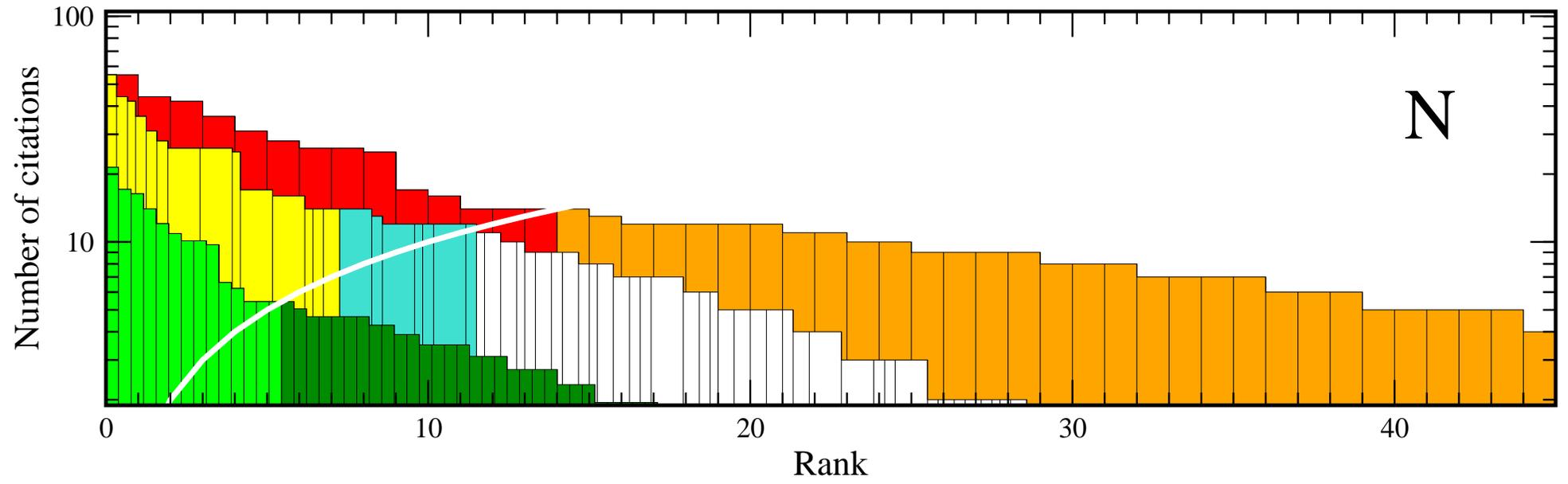

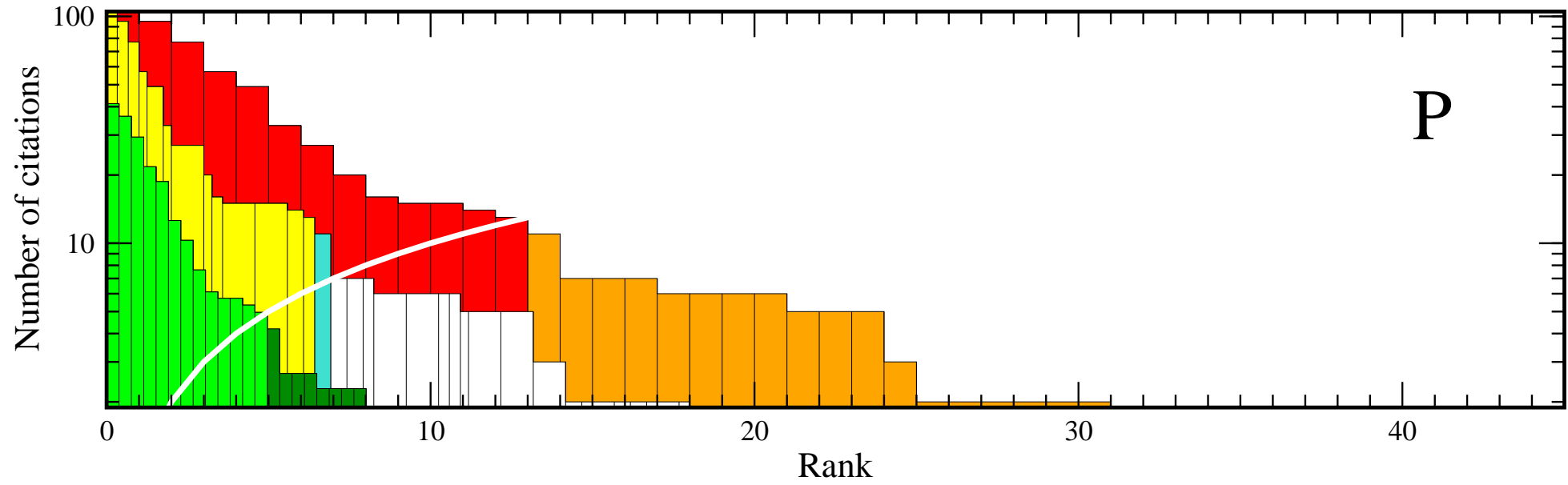

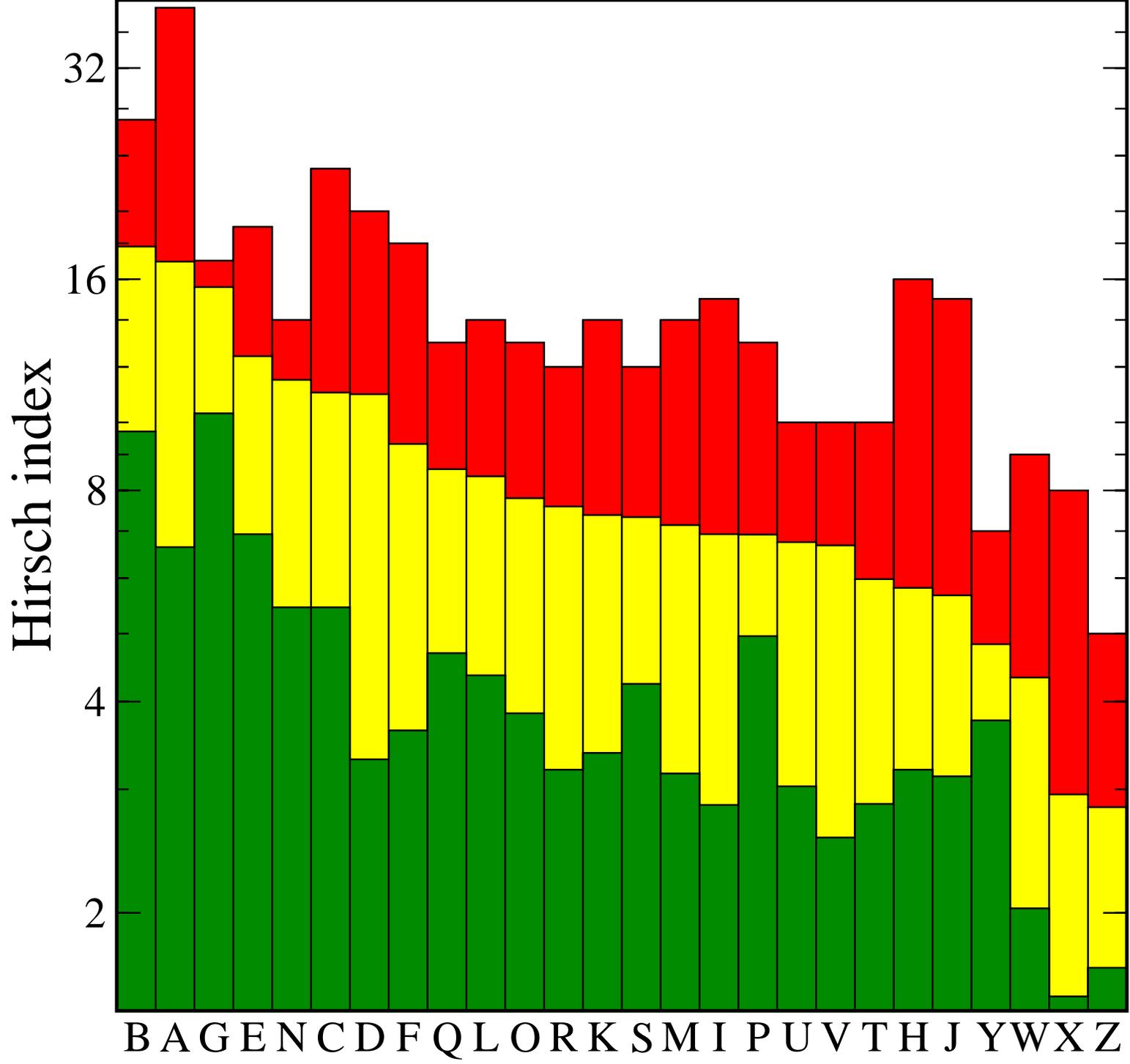

TABLE 1. The citation records for the 20 most cited papers in data sets V, W, X, and Y. Besides the number of citations $c(r)$ and the number of authors $a(r)$ the effective rank $r_{\text{eff}}(r)$ as defined in Equation (4) is given. For each case, the citation count of the last paper which contributes to the $h$-core, i.e. fulfilling Equation (1), is given in bold italics and the citation count of the last paper which enters the $h_{\text{m}}$-core, i.e., which fulfills Equation (5) is given in bold face.

| $r$ | $c^V(r)$ | $a^V(r)$ | $r_{\text{eff}}^V(r)$ | $c^W(r)$ | $a^W(r)$ | $r_{\text{eff}}^W(r)$ | $c^X(r)$ | $a^X(r)$ | $r_{\text{eff}}^X(r)$ | $c^Y(r)$ | $a^Y(r)$ | $r_{\text{eff}}^Y(r)$ |
|---|---|---|---|---|---|---|---|---|---|---|---|---|
| 1 | 79 | 3 | 0.33 | 42 | 4 | 0.25 | 204 | 8 | 0.13 | 19 | 2 | 0.50 |
| 2 | 34 | 4 | 0.58 | 21 | 3 | 0.58 | 15 | 3 | 0.46 | 14 | 2 | 1.00 |
| 3 | 32 | 4 | 0.83 | 16 | 4 | 0.83 | 14 | 6 | 0.63 | 11 | 2 | 1.50 |
| 4 | 25 | 2 | 1.33 | 12 | 8 | 0.96 | 11 | 3 | 0.96 | 10 | 2 | 2.00 |
| 5 | 16 | 4 | 1.58 | 12 | 3 | 1.29 | 10 | 8 | 1.08 | 9 | 3 | 2.33 |
| 6 | 13 | 4 | 1.83 | 10 | 6 | 1.46 | 10 | 4 | 1.33 | 7 | 1 | 3.33 |
| 7 | 12 | 10 | 1.93 | 9 | 3 | 1.79 | 9 | 5 | 1.53 | *7* | 1 | 4.33 |
| 8 | 11 | 2 | 2.43 | 9 | 5 | 1.99 | **8** | 5 | 1.73 | **6** | 2 | 4.83 |
| 9 | 11 | 3 | 2.77 | **9** | 4 | 2.24 | 8 | 5 | 1.93 | 5 | 1 | 5.83 |
| 10 | *11* | 3 | 3.10 | 8 | 3 | 2.58 | 7 | 8 | 2.06 | 5 | 4 | 6.08 |
| 11 | 8 | 3 | 3.43 | 8 | 3 | 2.91 | 7 | 7 | 2.20 | 4 | 3 | 6.42 |
| 12 | 8 | 1 | 4.43 | 8 | 6 | 3.08 | 6 | 6 | 2.37 | 4 | 1 | 7.42 |
| 13 | 8 | 2 | 4.93 | 8 | 4 | 3.33 | 5 | 4 | 2.62 | 4 | 1 | 8.42 |
| 14 | 8 | 4 | 5.18 | 7 | 3 | 3.66 | 4 | 6 | 2.78 | 3 | 1 | 9.42 |
| 15 | 7 | 2 | 5.68 | 7 | 8 | 3.78 | **3** | 5 | 2.98 | 2 | 2 | 9.92 |
| 16 | **7** | 1 | 6.68 | 7 | 8 | 3.91 | 3 | 6 | 3.15 | 2 | 2 | 10.42 |
| 17 | 6 | 2 | 7.18 | 5 | 4 | 4.16 | 3 | 6 | 3.32 | 2 | 1 | 11.42 |
| 18 | 6 | 2 | 7.68 | **5** | 6 | 4.33 | 3 | 6 | 3.48 | 1 | 1 | 12.42 |
| 19 | 5 | 2 | 8.18 | 4 | 3 | 4.66 | 3 | 6 | 3.65 | 1 | 2 | 12.92 |
| 20 | 5 | 3 | 8.52 | 4 | 5 | 4.86 | 2 | 5 | 3.85 | 0 | 2 | 13.42 |

TABLE 2. Hirsch index without and with taking multiple co-authorship into account as described in the text. Also given are the average number of $\bar{a}(h)$ of authors in the $h$-core, the compression $r_{\text{eff}}(h)$ of the $h$-core due to the fractionalised counting and the value $r(h_m)$, reflecting the size of the $h_m$-core. The last column shows the order in which the data sets appear after the list is sorted according to the modified index.

| data set | $h$ | $\bar{a}(h)$ | $h_I$ | $r_{\text{eff}}(h)$ | $r(h_m)$ | $h_m$ | $h_m/h$ | $\mathrm{O}[h_m]$ |
|---|---|---|---|---|---|---|---|---|
| A | 39 | 5.87 | 6.64 | 7.45 | 90 | 16.95 | 0.43 | 2 |
| B | 27 | 2.78 | 9.71 | 11.36 | 42 | 17.81 | 0.66 | 1 |
| C | 23 | 4.22 | 5.45 | 6.33 | 41 | 11.03 | 0.48 | 6 |
| D | 20 | 6.05 | 3.31 | 3.57 | 57 | 10.97 | 0.55 | 7 |
| E | 19 | 2.74 | 6.93 | 10.64 | 23 | 12.43 | 0.65 | 4 |
| F | 18 | 4.94 | 3.64 | 4.29 | 39 | 9.32 | 0.52 | 8 |
| G | 17 | 1.65 | 10.30 | 15.59 | 17 | 15.59 | 0.92 | 3 |
| H | 16 | 5.00 | 3.20 | 4.36 | 27 | 5.81 | 0.36 | 21 |
| I | 15 | 5.27 | 2.85 | 4.01 | 28 | 6.93 | 0.46 | 16 |
| J | 15 | 4.80 | 3.13 | 3.72 | 21 | 5.67 | 0.38 | 22 |
| K | 14 | 4.14 | 3.38 | 3.96 | 25 | 7.38 | 0.53 | 13 |
| L | 14 | 3.21 | 4.36 | 5.06 | 22 | 8.38 | 0.60 | 10 |
| M | 14 | 4.43 | 3.16 | 5.56 | 23 | 7.14 | 0.51 | 15 |
| N | 14 | 2.57 | 5.45 | 7.25 | 21 | 11.50 | 0.82 | 5 |
| O | 13 | 3.38 | 3.85 | 4.68 | 22 | 7.80 | 0.60 | 11 |
| P | 13 | 2.62 | 4.96 | 6.42 | 14 | 6.92 | 0.53 | 17 |
| Q | 13 | 2.77 | 4.69 | 6.00 | 20 | 8.58 | 0.66 | 9 |
| R | 12 | 3.75 | 3.20 | 4.81 | 17 | 7.59 | 0.63 | 12 |
| S | 12 | 2.83 | 4.24 | 4.92 | 17 | 7.33 | 0.61 | 14 |
| T | 10 | 3.50 | 2.86 | 3.03 | 21 | 5.98 | 0.60 | 20 |
| U | 10 | 3.30 | 3.03 | 3.58 | 18 | 6.75 | 0.68 | 18 |
| V | 10 | 3.90 | 2.56 | 3.10 | 16 | 6.68 | 0.67 | 19 |
| W | 9 | 4.44 | 2.03 | 2.24 | 18 | 4.33 | 0.48 | 24 |
| X | 8 | 5.25 | 1.52 | 1.73 | 15 | 2.98 | 0.37 | 25 |
| Y | 7 | 1.86 | 3.76 | 4.33 | 8 | 4.83 | 0.69 | 23 |
| Z | 5 | 3.00 | 1.67 | 1.83 | 6 | 2.83 | 0.57 | 26 |